\documentclass[12pt,a4paper]{article} 

\usepackage{amsfonts}
\usepackage{amssymb}
\usepackage{amsmath}
\usepackage[english]{babel}
\usepackage[cyr]{aeguill}
\usepackage{graphicx}
\usepackage[T1]{fontenc}
\usepackage{multirow} 
\usepackage{rotating}
\usepackage{textcomp}
\usepackage{verbatim} 
\usepackage{hyperref}
\hypersetup{pdfborder={0 0 0}}
\usepackage{pstricks}
\usepackage{colortbl}
\usepackage{setspace}
\usepackage{afterpage}
\usepackage{apacite}
\usepackage{caption}

\title{Inflated Discrete Beta Regression Models \\ for Likert and Discrete Rating Scale Outcomes}
\date{}
\author{ Cedric Taverne$^1$ \& Philippe Lambert$^{1,2}$ \\ {\footnotesize $^1$ Institut de Statistique, Biostatistique et Sciences Actuarielles,} \\ {\footnotesize Université Catholique de Louvain, Belgium} \\ {\footnotesize cedric.taverne@uclouvain.be} \\ {\footnotesize $^2$ Institut des Sciences Humaines et Sociales, Université de Liège, Belgium } \\ {\footnotesize p.lambert@ulg.ac.be} }

\definecolor{Gray}{gray}{0.9}

\begin{document} 


\doublespacing

\maketitle

\begin{abstract}
Discrete ordinal responses such as Likert scales are regularly proposed in questionnaires and used as dependent variable in modeling. The response distribution for such scales is always discrete, with bounded support and often skewed. In addition, one particular level of the scale is frequently inflated as it cumulates respondents who invariably choose that particular level (typically the middle or one extreme of the scale) without hesitation with those who chose that alternative but might have selected a neighboring one. The inflated discrete beta regression (IDBR) model addresses those four critical characteristics that have never been taken into account simultaneously by existing models. The mean and the dispersion of rates are jointly regressed on covariates using an underlying beta distribution. The probability that choosers of the inflated level invariably make that choice is also regressed on covariates. Simulation studies used to evaluate the statistical properties of the IDBR model suggest that it produces more precise predictions than competing models. The ability to jointly model the location and dispersion of (the distribution of) an ordinal response, as well as to characterize the profile of subject selecting an "inflated" alternative are the most relevant features of the IDBR model. It is illustrated with the analysis of the political positioning on a "left-right" scale of the Belgian respondents in the 2012 European Social Survey.
\end{abstract}


\let\thefootnote\relax\footnote{Requests for reprints should be sent to Cedric Taverne, Université catholique de Louvain, Louvain-la-Neuve, Belgium. E-mail: cedric.taverne@uclouvain.be}
\let\thefootnote\relax\footnote{The authors acknowledge financial support from IAP research network P7/06 of the Belgian Government (Belgian Science Policy), and from the contract "Projet d'Actions de Recherche Concertées" (ARC) 11/16-039 of the "Communautée française de Belgique", granted by the "Académie universitaire Louvain".}
\newpage 


{\centering \section{Introduction} }
\label{sec:Intro}


Whatever the subject of interest, discrete scales are used everywhere in questionnaires. It could be Likert scales from \textit{Strongly disagree} to \textit{Strongly agree}, rating scales from 0 to 10 or the "left-right" political scale such as for question B19 of the European Social Survey \cite{ESS6}. In applications, such data are generally analyzed using classical linear regression. Yet, this approach has several statistical limits: rates on such scales are always bounded, often skewed and frequently inflated at one of the bounds of the scale. Last but not least, the discreteness of the scale is not recognized in estimation and predictions when using that model. 

Among the alternatives to linear regression, discrete response models naturally address the discreteness of the scale and the existence of boundaries. Nevertheless, skewed and inflated responses are less effectively handled by ordered logit or probit regressions. The multinomial versions of those models perform better there but are heavily parametrized and complicated to interpret. Another alternative to linear regression are the various versions of beta regression proposed in the literature for modeling rates and proportions (e.g., \citeNP{BrehmGates1993, paolino2001, FerrariCribariNeto2004, BranscumJohnson2007}; Simas, Barreto-Souza, \& Rocha, \citeyearNP{ SimasBarretoSouza2010}). The beta distribution is bounded and can be skewed but makes the assumption of continuity. As far as we know, none of these methods has been adapted to discrete scales and their quality on such scales has never been studied. Inflated beta regressions also exist for continuous outcomes \cite<e.g.,>{OspinaFerrari2010, WieczorekHawala2011, OspinaFerrari2012}. Nevertheless, the inflation is considered outside the support of the beta distribution which is not suitable to deal with discrete scales. 

The model proposed here addresses the four expounded limits by adapting the beta regression proposed by \citeA{SimasBarretoSouza2010}. The discrete scale is assumed to be the observed counterpart of an underlying beta distribution where the mean and the dispersion are jointly regressed on covariates. The likelihood function is adapted to suit the discreteness of the scale. Skewness and boundedness are intrinsic features of the beta distribution. Finally, when one of the points of the support is inflated, a switch to an inflated discrete beta regression is made by jointly modeling an additional mass of probability on the inflated point that can be anywhere on the scale. 

The next section contains the theoretical description of the discrete beta regression model, while section 3 
describes the required amendments to deal with inflation. The fourth section is related to the estimation procedure and contains simulation studies of its properties. The fifth section compares the quality of predictions of the inflated discrete beta regression model with those of competing models. Finally, the sixth section illustrates our model on one question of the European Social Survey for which the inflated point is in the middle of an odd-level scale.


{\centering \section{Discrete Beta Regression}}
\label{sec:DBR}


Let $y^\star_i$ be the choice of individual $i$ on a $K$-level discrete scale $Y^\star$. This scale is considered as a random variable with unknown distribution but known support, ${\cal Y}^\star = \left\{a,a+h^\star,...,b-h^\star,b\right\}$: thus, we assume that each point on this scale are equally spaced and possibly labeled ("Strongly agree", etc.). In order to match the discrete scale with the support of the beta distribution, its support is rescaled into the unit interval using
\begin{equation} 
\mathbf{y} = \frac{\mathbf{y}^\star-a+h^\star}{b-a+h^\star} ~. \nonumber 
\end{equation} 
One gets a reduced $K$-level discrete scale $Y$ with rescaled support, ${\cal Y} = \left\{h,2h,...,1-h,1\right\}$ where $h = 1/K$. 

\begin{figure}[ht!]
\begin{center}
	\includegraphics{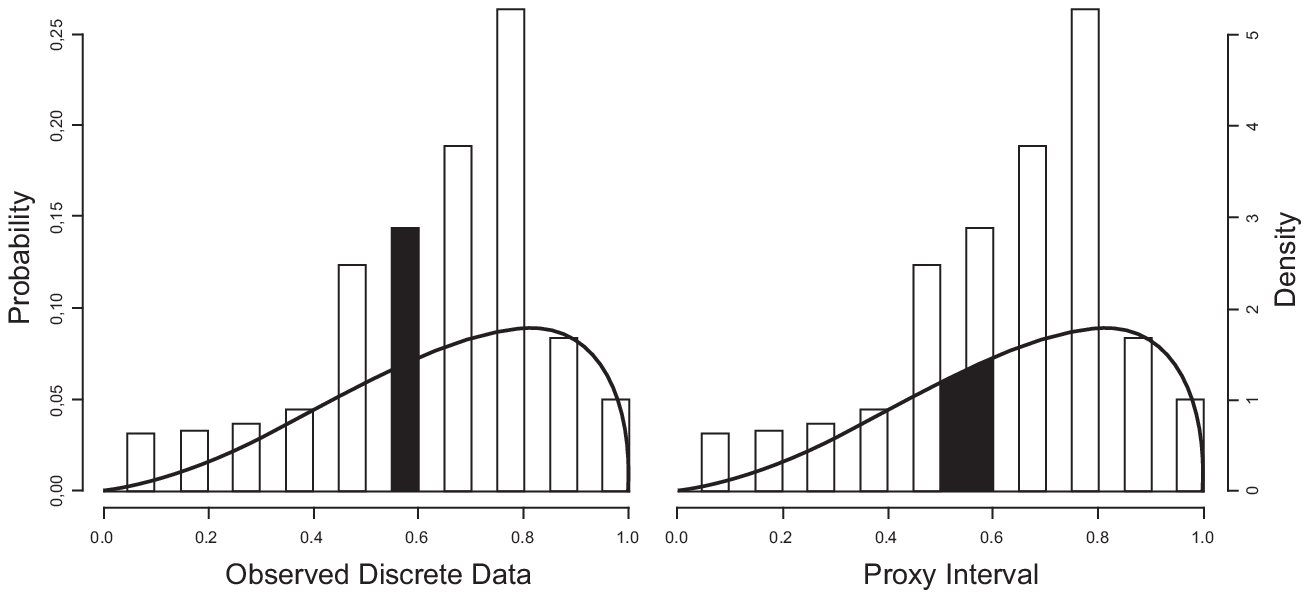}
	\caption{Observed discrete data and proxy interval considered under the beta density as described in equation (\ref{eq:DiscreteToContinousProbaNoInfl}).}
	\label{fig:ObsDataAndProxiArea}
\end{center}
\end{figure}

The discrete beta regression model is defined in two steps. Firstly, the rescaled discrete scale is linked to the continuous beta distribution using
\begin{equation}
P(Y = y) = P\left(y-h < U \le y \right) ~, \label{eq:DiscreteToContinousProbaNoInfl} 
\end{equation}
where $U \sim Be(p,q)$ with $p>0$, $q>0$ and $u \in (0,1)$. Figure \ref{fig:ObsDataAndProxiArea} illustrates the idea behind this link: the black area under the curve of the beta density is used as a proxy for the probability mass in the bar. 

Secondly, the parameters of the beta distribution are regressed on covariates. It is based on an alternative parametrization of the distribution enabling to model location and dispersion separately. The location parameter $\mu$ is the mean of the beta distribution given by 
\begin{equation}
 E(U) = \frac{p}{p+q} = \mu ~. \nonumber 
\end{equation}
The dispersion parameter $\phi$ is proportional to the variance of the distribution,
\begin{equation}
Var(U) = \frac{ p ~q }{ (p+q)^2(p+q+1) } =  \frac{\mu(1-\mu)}{p+q+1} = \mu(1-\mu)\phi ~. \nonumber 
\end{equation}
Both quantities take values in $\left(0,1\right)$. Link functions $\mathrm{g}_1$ and $\mathrm{g}_2$ are used to relate $\mu$ and $\phi$ to covariates,
\begin{eqnarray}
\mathrm{g}_1\left(\boldsymbol\mu\right) = \mathrm{f}_1(\mathbf{\mathrm{X}},\boldsymbol\beta) ~, & \qquad & \mathrm{g}_1: (0,1) \rightarrow \mathbb{R} ~, \label{eq:ExtBetaRegrModelMuGeneral} \\
\mathrm{g}_2\left(\boldsymbol\phi\right) = \mathrm{f}_2(\mathbf{\mathrm{Z}},\boldsymbol\theta) ~, & \qquad & \mathrm{g}_2: (0,1) \rightarrow \mathbb{R} ~. \label{eq:ExtBetaRegrModelPhiGeneral}
\end{eqnarray}
They are assumed to be strictly monotonic and twice differentiable. Logit links will be used in applications. The utility functions $\mathrm{f}_1$ and $\mathrm{f}_2$ will be assumed linear in this paper but this is not compulsory. We assume that the matrices $\partial \mathrm{f}_1(\mathbf{\mathrm{X}},\boldsymbol\beta) / \partial \boldsymbol\beta$ and $\partial \mathrm{f}_2(\mathbf{\mathrm{Z}},\boldsymbol\theta) / \partial \boldsymbol\theta$ have full rank. The covariates matrices $\mathbf{\mathrm{X}}$ and $\mathbf{\mathrm{Z}}$ could be (partially) the same but this is not compulsory.

The contribution of the $i^\mathrm{th}$ observation to the likelihood in the discrete beta regression is given by
\begin{equation}
P(Y_i = y_i\left|\mathbf{x}_i,\mathbf{z}_i,\boldsymbol\beta,\boldsymbol\theta\right.) = \frac{  \int_{y_i-h}^{y_i}{u^{\mathrm{p}(\mathbf{x}_i,\mathbf{z}_i,\boldsymbol\beta,\boldsymbol\theta)-1}(1-u)^{\mathrm{q}(\mathbf{x}_i,\mathbf{z}_i,\boldsymbol\beta,\boldsymbol\theta)-1} du}}{\mathrm{B}\left[\mathrm{p}(\mathbf{x}_i,\mathbf{z}_i,\boldsymbol\beta,\boldsymbol\theta)~,~\mathrm{q}(\mathbf{x}_i,\mathbf{z}_i,\boldsymbol\beta,\boldsymbol\theta)\right]} ~,  \label{eq:ExtBetaRegrLikeIndNoInflGeneral} 
\end{equation}
where $y_i$ is the observed point on the reduced scale, $\mathrm{p}(\mathbf{x}_i,\mathbf{z}_i,\boldsymbol\beta,\boldsymbol\theta) = \mathrm{g}_1^{-1}(\mathrm{f}_1(\mathbf{x}_i,$ $\boldsymbol\beta)) \left[\frac{1}{\mathrm{g}_2^{-1}(\mathrm{f}_2(\mathbf{z}_i,\boldsymbol\theta))}-1\right]$, $\mathrm{q}(\mathbf{x}_i,\mathbf{z}_i,\boldsymbol\beta,\boldsymbol\theta) = \left[1-\mathrm{g}_1^{-1}(\mathrm{f}_1(\mathbf{x}_i,\boldsymbol\beta))\right] \left[\frac{1}{\mathrm{g}_2^{-1}(\mathrm{f}_2(\mathbf{z}_i,\boldsymbol\theta))}-1\right]$ \\ and $\mathrm{B}\left[.,.\right]$ is the beta function. When logit links and linear utility functions are used, this general class of discrete beta regression can be written with $\mathrm{p}(\mathbf{x}_i,\mathbf{z}_i,\boldsymbol\beta,\boldsymbol\theta)=\frac{\exp\left(-\mathbf{z}_{i}'\boldsymbol\theta\right)}{1+\exp\left(-\mathbf{x}_{i}'\boldsymbol\beta\right)}$ and $\mathrm{q}(\mathbf{x}_i,\mathbf{z}_i,\boldsymbol\beta,\boldsymbol\theta)=\frac{\exp\left(-\mathbf{z}_{i}'\boldsymbol\theta\right)}{1+\exp\left(\mathbf{x}_{i}'\boldsymbol\beta\right)}$.

In contrast to the continuous beta regression model proposed by Simas et al. \citeyear{SimasBarretoSouza2010}, the discreteness of the scale is embodied in the model formulation. A parametrization of the beta distribution in term of dispersion is also preferred as consulting experience suggests that this concept is generally easier to explain to non-statisticians. Nevertheless, since the Simas' precision parameter is defined by $\varphi=p+q$, one has $\phi = \frac{1}{\varphi+1}$. Thus, if one uses a logarithm link in Simas' model and a logit link  in (\ref{eq:ExtBetaRegrModelPhiGeneral}) with linear utility functions on both sides, then the $\theta$'s are simply the opposite of Simas' precision parameters. Consequently, users can easily switch from dispersion to precision by changing the signs of the parameters.


{\centering \section{Inflated Discrete Beta Regression} }
\label{sec:IDBR}


In discrete ordinal responses, an unusual inflation in the distribution of the responses is regularly observed at one of the point of the scale's support, say the $k^\mathrm{th}$ one where $y=kh$ with $k \in \left\{1,2,...,K\right\}$ and $h=1/K$. It might be one of the bounds of the scales or the central level of odd-level scales. In such situation, one can suspect that choosers of that point are composed of two types of respondents: those who invariably choose that particular level without hesitation and those who choose that alternative but might have chosen the neighboring points on the scale as well. In the following, the probability related to the former is noted $\pi$ and is regressed on covariates whereas the behavior of the latter is described by the discrete beta regression (DBR) model defined in section 2. 

Both steps of modeling described before have to be adapted. Thus, one has
\begin{equation}
P(Y = y) = I(y=kh) ~ \pi ~+~ P\left(y-h < U \le y \right) ~ \left[1-\pi\right] ~,  \nonumber \end{equation}
where $U$ denotes the latent beta of the DBR component and $I(.)$ is an indicator function. The probability $\pi$ can also be related to covariates using
\begin{equation}
\mathrm{g}_0\left( \boldsymbol\pi \right) = \mathrm{f}_0(\mathbf{\mathrm{W}},\boldsymbol\gamma)~, \qquad \qquad \mathrm{g}_0: (0,1) \rightarrow \mathbb{R} ~, \label{eq:ExtBetaRegrModelInflGeneral}
\end{equation}
where $\mathrm{g}_0$ and $\mathrm{f}_0$ denote the link and utility functions respectively.

The contribution of the $i^\mathrm{th}$ observation to the likelihood of the inflated discrete beta regression (IDBR) is 
\begin{align}
& P(Y_i = y_i \left|\mathbf{x}_i,\mathbf{z}_i,\mathbf{w}_i,\boldsymbol\beta,\boldsymbol\theta,\boldsymbol\gamma\right.) = I(y_i=kh) ~	\mathrm{g}_0^{-1}\left(\mathrm{f}_0\left(\mathbf{w}_i,\boldsymbol\gamma\right)\right) \phantom{\frac{u^{p}}{\left[\mathbf{x}_i\right]}} \nonumber \\
& \qquad ~+~ \frac{  \int_{y_i-h}^{y_i}{u^{\mathrm{p}(\mathbf{x}_i,\mathbf{z}_i,\boldsymbol\beta,\boldsymbol\theta)-1}~(1-u)^{\mathrm{q}(\mathbf{x}_i,\mathbf{z}_i,\boldsymbol\beta,\boldsymbol\theta)-1} du}}{\mathrm{B}\left[\mathrm{p}(\mathbf{x}_i,\mathbf{z}_i,\boldsymbol\beta,\boldsymbol\theta)~,~\mathrm{q}(\mathbf{x}_i,\mathbf{z}_i,\boldsymbol\beta,\boldsymbol\theta)\right]} \times \left[1-\mathrm{g}_0^{-1}\left(\mathrm{f}_0\left(\mathbf{w}_i,\boldsymbol\gamma\right)\right)\right]    ~, \nonumber 
\end{align}
where elements of the DBR remain defined as in equation (\ref{eq:ExtBetaRegrLikeIndNoInflGeneral}). This general class of inflated discrete beta regression can be written as $\mathrm{g}_0^{-1}\left(\mathrm{f}_0\left(\mathbf{w}_i,\boldsymbol\gamma\right)\right) = \frac{1}{1+\exp\left(-\mathbf{w}_i'\boldsymbol\gamma\right)}$ when logit link and linear utility function are used. 

In order to determine if an inflation point must be included in the model, users are advised to visualize the empirical distribution of the response within classes of potential covariates and to wonder if the choice of the inflated point on the scale might correspond to a specific behavior or pattern of respondents for whom the proposition is \textit{clearly not an option} or is a \textit{systematic choice}. A post-modeling comparison of the DBR and IDBR models can also be conducted using the Bayes factor or other criteria.

Zero-inflated, one-inflated and zero-and-one-inflated beta regression model already exist for continuous outcomes (\citeNP{OspinaFerrari2010}; Wieczorek \& Hawala, \citeyearNP{WieczorekHawala2011}; \citeNP{OspinaFerrari2012}). Contrary to the IDBR, those models do not take into account the discreteness of the scale and are based on a disjoint support: the inflation is observed outside the support of the beta distribution. The IDBR is more comparable to the zero-inflated Poisson regression proposed by \citeA{Lambert1992} where 0 is contained in the support of the Poisson distribution. Furthermore, the inflation can be set anywhere (where found appropriate) on the scale in the IDBR model. This will be illustrated in the case study in section 6.


{\centering \section{Estimation} }
\label{sec:ExtBetaRegrEstimate}


A Bayesian estimation of the model can be obtained by sampling its joint posterior using a Metropolis algorithm with adaptive scale parameters \cite{AtchadeRosenthal2005}. For this purpose, the authors developed a Fortran code called from R. Since there is no theoretical constraint on the regression parameters, non informative uniform priors were considered. With logit links for $\mathrm{g}_0$, $\mathrm{g}_1$ and $\mathrm{g}_2$, the values of regression parameters for standardized covariates are just constrained to take values in $\left[-10;10\right]$ since changes outside this interval do not affect the first four decimal places of $\pi$, $\mu$ and $\phi$, see (\ref{eq:ExtBetaRegrModelMuGeneral}), (\ref{eq:ExtBetaRegrModelPhiGeneral}) and (\ref{eq:ExtBetaRegrModelInflGeneral}) respectively. After a burn in period of 1000 iterations, convergence is monitored using the Gelman diagnostic tool \cite{GelmanRubin1992} on 3 parallel Markov chains Monte Carlo (MCMC) of length 1000 each.  The initial values for the regression parameters in the first chain were selected using marginal mean, variance and proportion estimates:
\begin{eqnarray}
\hat{\gamma}_{0,\mathrm{init}_1} & = & \mathrm{g}_0\left( \frac{1}{n} \sum_{i=1}^n{ I\left(y_i=kh\right) } \right) ~, \nonumber \\ 
\hat{\beta}_{0,\mathrm{init}_1} & = & \mathrm{g}_1\left( ~\bar{y}~ \right) ~, \phantom{\frac{1}{n}}  \nonumber \\ 
\hat{\theta}_{0,\mathrm{init}_1} & = & \mathrm{g}_2\left( \frac{\frac{1}{n-1} \sum_{i=1}^n{ \left(y_i-\bar{y}\right)^2 } }{ \bar{y} \left(1-\bar{y}\right) } \right) ~, \nonumber 
\end{eqnarray}
where $\bar{y} = \frac{1}{n} \sum_{i=1}^n{ y_i }$ and $n$ is the sample size. The idea is to set those intercepts to values that approximate the marginal distribution of the response. The other parameters are set to 0.

The initial values for the second chain approximately correspond to an uniform distribution for a response on $(0,1)$ (and, thus, with mean $1/2$ and variance $1/12$) showing no inflation:
\begin{eqnarray}
	\hat{\gamma}_{0,\mathrm{init}_2} & = & -9 ~ \approx ~ \mathrm{g}_1\left(0\right) ~, \nonumber \\ 
	\hat{\beta}_{0,\mathrm{init}_2} & = & \mathrm{g}_1\left(1/2\right) ~, \nonumber \\ 
	\hat{\theta}_{0,\mathrm{init}_2} & = & \mathrm{g}_2\left(1/3\right) ~. \nonumber  
\end{eqnarray}

For the third chain, the initial values of the parameters were estimated using the continuous beta regression of \citeA{SimasBarretoSouza2010} for location and dispersion and the frequentist estimation of a binary logit model for the inflation.

The median of the 3000 draws following the burn in is used as a point estimate for each parameter and to evaluate the bias, the standard deviation and the root mean squared error in simulations. The credibility intervals considered are the highest posterior density (HPD) intervals calculated using the function \texttt{HPDinterval} in the R package \texttt{coda} (Plummer, Best, Cowles, \& Vines, \citeyearNP{PlummerBest2006}).

The quality of our estimation scheme was evaluated in details under various simulation settings: inflated and non inflated scales, varying sample sizes, high correlations between the covariates and omission of significant explanatory variables. Within those simulations, the acceptance rates, the effective sample sizes and the convergence diagnostics were considered as well as the bias and the precision of parameter estimates. The covering rates of HPD intervals and their length were monitored too. Tables \ref{tab:ExtBetaRegrSimuSampleSize6} and \ref{tab:ExtBetaRegrSimuSampleSize11} provide a selected summary of those simulation results; detailed results are available on request. 

\begin{table}[ht!]
\scriptsize \centering
\caption{Sensitivity of the estimation scheme to various sample sizes on a 6-level inflated discrete scale with 4 continuous (V1 to V4) and 3 dummy (D1 to D3) covariates across 500 simulations}
\label{tab:ExtBetaRegrSimuSampleSize6}
\begin{tabular}{lcrrrrrr}
  \hline
 & Sample & True & \multirow{2}{*}{Bias} & \multicolumn{1}{c}{Emp.} & \multicolumn{1}{c}{\multirow{2}{*}{RMSE}} & \multicolumn{2}{c}{HPD(95\%)}  \\ 
 & Size & \multicolumn{1}{c}{Param.} &  & \multicolumn{1}{c}{SD} & & Cov. & Length  \\ 
  \hline
\multicolumn{3}{l}{\textbf{Inflation submodel}}  &  &  &  &  &    \\ 
\arrayrulecolor{Gray}  \hline
\multirow{2}{*}{Int.} & n=300 & \multirow{2}{*}{-4.500} & -0.896 & 2.084 & 2.267 & 0.898 & 6.747  \\ 
 & n=900 &  & -0.241 & 0.901 & 0.931 & 0.942 & 3.394 \\ 
\hline
  \multirow{2}{*}{V1} & n=300 & \multirow{2}{*}{1.000} & 0.126 & 0.321 & 0.344 & 0.902 & 1.088 \\ 
 & n=900 &  & 0.039 & 0.145 & 0.150 & 0.928 & 0.547  \\ 
\hline
  \multirow{2}{*}{V2} & n=300 & \multirow{2}{*}{0.000} & 0.111 & 0.306 & 0.325 & 0.930 & 1.122  \\ 
 & n=900 &  & 0.023 & 0.148 & 0.150 & 0.942 & 0.580  \\ 
\hline
  \multirow{2}{*}{V3} & n=300 & \multirow{2}{*}{0.300} & 0.015 & 0.255 & 0.255 & 0.944 & 0.962  \\ 
 & n=900 &  & 0.011 & 0.120 & 0.120 & 0.964 & 0.491  \\ 
\hline
  \multirow{2}{*}{V4} & n=300 & \multirow{2}{*}{-0.500} & -0.101 & 0.300 & 0.317 & 0.902 & 1.027   \\ 
 & n=900 &  & -0.027 & 0.143 & 0.145 & 0.920 & 0.520  \\ 
\hline
  \multirow{2}{*}{D1} & n=300 & \multirow{2}{*}{-0.500} & -0.061 & 0.509 & 0.512 & 0.950 & 1.940 \\ 
 & n=900 &  & -0.052 & 0.261 & 0.266 & 0.942 & 0.985  \\ 
\hline
  \multirow{2}{*}{D2} & n=300 & \multirow{2}{*}{0.000} & 0.076 & 0.533 & 0.538 & 0.946 & 1.949  \\ 
 & n=900 &  & 0.005 & 0.258 & 0.258 & 0.936 & 0.986  \\ 
\hline
  \multirow{2}{*}{D3} & n=300 & \multirow{2}{*}{0.000} & -0.022 & 0.557 & 0.557 & 0.920 & 1.908  \\ 
 & n=900 &  & -0.006 & 0.253 & 0.253 & 0.960 & 0.970  \\ 
\arrayrulecolor{black}  \hline
\multicolumn{3}{l}{\textbf{Location submodel}}  &  &  &  &  &    \\ 
\arrayrulecolor{Gray}  \hline
  \multirow{2}{*}{Int.} & n=300 & \multirow{2}{*}{-1.000} & -0.013 & 0.249 & 0.249 & 0.932 & 0.931   \\ 
 & n=900 &  & -0.012 & 0.139 & 0.139 & 0.942 & 0.527  \\ 
\hline
  \multirow{2}{*}{V1} & n=300 & \multirow{2}{*}{-0.200} & -0.005 & 0.042 & 0.042 & 0.928 & 0.155  \\ 
 & n=900 &  & -0.000 & 0.022 & 0.022 & 0.948 & 0.087   \\ 
\hline
  \multirow{2}{*}{V2} & n=300 & \multirow{2}{*}{0.900} & 0.010 & 0.050 & 0.050 & 0.920 & 0.183  \\ 
 & n=900 &  & 0.003 & 0.027 & 0.027 & 0.946 & 0.104  \\ 
\hline
  \multirow{2}{*}{V3} & n=300 & \multirow{2}{*}{0.000} & -0.004 & 0.040 & 0.040 & 0.932 & 0.149  \\ 
 & n=900 &  & -0.001 & 0.022 & 0.022 & 0.940 & 0.084  \\ 
\hline
  \multirow{2}{*}{V4} & n=300 & \multirow{2}{*}{-0.400} & -0.001 & 0.043 & 0.043 & 0.928 & 0.156   \\ 
 & n=900 &  & 0.001 & 0.023 & 0.023 & 0.942 & 0.088  \\ 
\hline
  \multirow{2}{*}{D1} & n=300 & \multirow{2}{*}{0.000} & 0.007 & 0.077 & 0.078 & 0.942 & 0.292  \\ 
 & n=900 &  & -0.000 & 0.043 & 0.043 & 0.936 & 0.166  \\ 
\hline
  \multirow{2}{*}{D2} & n=300 & \multirow{2}{*}{0.700} & 0.003 & 0.080 & 0.080 & 0.946 & 0.298  \\ 
 & n=900 &  & 0.005 & 0.044 & 0.044 & 0.940 & 0.169  \\ 
\hline
  \multirow{2}{*}{D3} & n=300 & \multirow{2}{*}{0.000} & 0.002 & 0.073 & 0.073 & 0.958 & 0.298  \\ 
 & n=900 &  & -0.002 & 0.044 & 0.044 & 0.940 & 0.169  \\ 
\arrayrulecolor{black}  \hline
\multicolumn{3}{l}{\textbf{Dispersion submodel}}  &  &  &  &  &   \\ 
\arrayrulecolor{Gray}  \hline
  \multirow{2}{*}{Int.} & n=300 & \multirow{2}{*}{-3.000} & -0.075 & 0.891 & 0.893 & 0.936 & 3.322   \\ 
 & n=900 &  & 0.001 & 0.465 & 0.465 & 0.950 & 1.767   \\ 
\hline
  \multirow{2}{*}{V1} & n=300 & \multirow{2}{*}{0.000} & 0.015 & 0.151 & 0.151 & 0.938 & 0.550  \\ 
 & n=900 &  & 0.002 & 0.073 & 0.073 & 0.940 & 0.293  \\ 
\hline
  \multirow{2}{*}{V2} & n=300 & \multirow{2}{*}{-0.200} & -0.004 & 0.172 & 0.171 & 0.938 & 0.623  \\ 
 & n=900 &  & -0.003 & 0.085 & 0.085 & 0.948 & 0.332  \\ 
\hline
  \multirow{2}{*}{V3} & n=300 & \multirow{2}{*}{0.400} & 0.031 & 0.139 & 0.142 & 0.932 & 0.533  \\ 
 & n=900 &  & 0.010 & 0.077 & 0.077 & 0.932 & 0.285  \\ 
\hline
  \multirow{2}{*}{V4} & n=300 & \multirow{2}{*}{-0.200} & -0.024 & 0.150 & 0.152 & 0.932 & 0.554   \\ 
 & n=900 &  & -0.010 & 0.077 & 0.078 & 0.936 & 0.293  \\ 
\hline
  \multirow{2}{*}{D1} & n=300 & \multirow{2}{*}{0.000} & -0.015 & 0.274 & 0.274 & 0.932 & 1.023  \\ 
 & n=900 &  & 0.002 & 0.142 & 0.142 & 0.936 & 0.552  \\ 
\hline
  \multirow{2}{*}{D2} & n=300 & \multirow{2}{*}{0.000} & 0.011 & 0.280 & 0.280 & 0.950 & 1.036  \\ 
 & n=900 &  & -0.004 & 0.140 & 0.140 & 0.960 & 0.560  \\ 
\hline
  \multirow{2}{*}{D3} & n=300 & \multirow{2}{*}{0.500} & 0.029 & 0.279 & 0.280 & 0.934 & 1.023  \\ 
 & n=900 &  & 0.009 & 0.145 & 0.145 & 0.938 & 0.553  \\ 
\arrayrulecolor{black} \hline
\end{tabular}
\end{table}

\begin{table}[ht!]
\scriptsize \centering
\begin{center}
\caption{Sensitivity of the estimation scheme to various sample sizes on a 11-level inflated discrete scale with 4 continuous (V1 to V4) and 3 dummy (D1 to D3) covariates across 500 simulations}
\label{tab:ExtBetaRegrSimuSampleSize11}
\begin{tabular}{lcrrrrrr}
  \hline
 & Sample & \multicolumn{1}{c}{True} & \multicolumn{1}{c}{\multirow{2}{*}{Bias}} & \multicolumn{1}{c}{Emp.} & \multirow{2}{*}{RMSE} & \multicolumn{2}{c}{HPD(95\%)}  \\ 
 & Size & \multicolumn{1}{c}{Param.} &  & \multicolumn{1}{c}{SD} & & Cov. & Length \\ 
  \hline
\multicolumn{3}{l}{\textbf{Inflation submodel}}  &  &  &  &  &  \\ 
\arrayrulecolor{Gray}  \hline
\multirow{2}{*}{Int.} & n=300 & \multirow{2}{*}{-5.000} & -0.805 & 2.052 & 2.202 & 0.928 & 7.306   \\ 
    & n=900 &  & -0.252 & 0.936 & 0.968 & 0.946 & 3.722  \\ 
\hline
 \multirow{2}{*}{V1} & n=300 & \multirow{2}{*}{1.000} & 0.133 & 0.334 & 0.359 & 0.922 & 1.171  \\ 
     & n=900 &  & 0.031 & 0.161 & 0.164 & 0.932 & 0.595   \\ 
\hline
 \multirow{2}{*}{V2} & n=300 & \multirow{2}{*}{0.000} & 0.056 & 0.330 & 0.334 & 0.930 & 1.173  \\ 
      & n=900 &  & 0.021 & 0.156 & 0.158 & 0.940 & 0.608  \\ 
\hline
 \multirow{2}{*}{V3} & n=300 & \multirow{2}{*}{0.300} & 0.004 & 0.289 & 0.288 & 0.940 & 1.052   \\ 
       & n=900 &  & 0.014 & 0.144 & 0.145 & 0.938 & 0.544  \\ 
\hline
 \multirow{2}{*}{V4} & n=300 & \multirow{2}{*}{-0.500} & -0.085 & 0.318 & 0.329 & 0.920 & 1.102  \\ 
        & n=900 &  & -0.024 & 0.150 & 0.151 & 0.924 & 0.565  \\ 
\hline
 \multirow{2}{*}{D1} & n=300 & \multirow{2}{*}{-0.500} & -0.078 & 0.592 & 0.597 & 0.952 & 2.157  \\ 
         & n=900 &  & -0.031 & 0.280 & 0.281 & 0.944 & 1.100   \\ 
\hline
 \multirow{2}{*}{D2} & n=300 & \multirow{2}{*}{0.000} & 0.019 & 0.591 & 0.591 & 0.958 & 2.134  \\ 
          & n=900 &  & -0.001 & 0.280 & 0.280 & 0.952 & 1.087  \\ 
\hline
 \multirow{2}{*}{D3} & n=300 & \multirow{2}{*}{0.000} & -0.006 & 0.592 & 0.592 & 0.942 & 2.120  \\ 
           & n=900 &  & -0.007 & 0.299 & 0.299 & 0.936 & 1.080  \\ 
\arrayrulecolor{black}  \hline
\multicolumn{3}{l}{\textbf{Location submodel}}  &  &  &  &  &  \\ 
\arrayrulecolor{Gray}  \hline
 \multirow{2}{*}{Int.} & n=300 & \multirow{2}{*}{-1.000} & -0.002 & 0.209 & 0.209 & 0.948 & 0.823  \\ 
            & n=900 &  & 0.006 & 0.112 & 0.112 & 0.962 & 0.464   \\ 
\hline
 \multirow{2}{*}{V1} & n=300 & \multirow{2}{*}{-0.200} & -0.003 & 0.037 & 0.037 & 0.938 & 0.134   \\ 
             & n=900 &  & -0.002 & 0.019 & 0.019 & 0.960 & 0.075  \\ 
\hline
 \multirow{2}{*}{V2} & n=300 & \multirow{2}{*}{0.900} & 0.007 & 0.042 & 0.042 & 0.936 & 0.154  \\ 
              & n=900 &  & 0.003 & 0.022 & 0.022 & 0.950 & 0.087   \\ 
\hline
 \multirow{2}{*}{V3} & n=300 & \multirow{2}{*}{0.000} & -0.003 & 0.034 & 0.034 & 0.934 & 0.131  \\ 
               & n=900 &  & -0.002 & 0.020 & 0.020 & 0.932 & 0.073  \\ 
\hline
 \multirow{2}{*}{V4} & n=300 & \multirow{2}{*}{-0.400} & -0.002 & 0.036 & 0.036 & 0.930 & 0.136   \\ 
                & n=900 &  & -0.002 & 0.020 & 0.020 & 0.948 & 0.077   \\ 
\hline
 \multirow{2}{*}{D1} & n=300 & \multirow{2}{*}{0.000} & -0.001 & 0.069 & 0.069 & 0.934 & 0.260  \\ 
                 & n=900 &  & -0.001 & 0.038 & 0.038 & 0.932 & 0.146  \\ 
\hline
 \multirow{2}{*}{D2} & n=300 & \multirow{2}{*}{0.700} & 0.005 & 0.071 & 0.071 & 0.926 & 0.262   \\ 
                  & n=900 &  & 0.001 & 0.039 & 0.039 & 0.938 & 0.148   \\ 
\hline
 \multirow{2}{*}{D3} & n=300 & \multirow{2}{*}{0.000} & -0.002 & 0.068 & 0.068 & 0.938 & 0.265 \\ 
                   & n=900 &  & 0.000 & 0.040 & 0.040 & 0.936 & 0.149  \\ 
\arrayrulecolor{black}  \hline
\multicolumn{3}{l}{\textbf{Dispersion submodel}}  &  &  &  &  &   \\ 
\arrayrulecolor{Gray}  \hline
 \multirow{2}{*}{Int.} & n=300 & \multirow{2}{*}{-3.000} & 0.031 & 0.671 & 0.671 & 0.948 & 2.561  \\ 
                    & n=900 &  & 0.007 & 0.361 & 0.360 & 0.948 & 1.402  \\ 
\hline
 \multirow{2}{*}{V1} & n=300 & \multirow{2}{*}{0.000} & 0.006 & 0.115 & 0.115 & 0.934 & 0.421  \\ 
                     & n=900 &  & 0.002 & 0.061 & 0.061 & 0.934 & 0.229  \\ 
\hline
 \multirow{2}{*}{V2} & n=300 & \multirow{2}{*}{-0.200} & -0.011 & 0.118 & 0.118 & 0.944 & 0.451  \\ 
                      & n=900 &  & -0.006 & 0.063 & 0.063 & 0.940 & 0.251   \\ 
\hline
 \multirow{2}{*}{V3} & n=300 & \multirow{2}{*}{0.400} & 0.009 & 0.106 & 0.107 & 0.942 & 0.410   \\ 
                       & n=900 &  & 0.001 & 0.057 & 0.057 & 0.948 & 0.224  \\ 
\hline
 \multirow{2}{*}{V4} & n=300 & \multirow{2}{*}{-0.200} & -0.011 & 0.105 & 0.106 & 0.954 & 0.423  \\ 
                        & n=900 &  & -0.001 & 0.057 & 0.057 & 0.944 & 0.229  \\ 
\hline
 \multirow{2}{*}{D1} & n=300 & \multirow{2}{*}{0.000} & -0.003 & 0.214 & 0.213 & 0.934 & 0.801  \\ 
                         & n=900 &  & 0.002 & 0.116 & 0.116 & 0.942 & 0.439  \\ 
\hline
 \multirow{2}{*}{D2} & n=300 & \multirow{2}{*}{0.000} & -0.006 & 0.206 & 0.206 & 0.956 & 0.803  \\ 
                          & n=900 &  & 0.004 & 0.117 & 0.117 & 0.932 & 0.443  \\ 
\hline
 \multirow{2}{*}{D3} & n=300 & \multirow{2}{*}{0.500} & -0.003 & 0.214 & 0.214 & 0.944 & 0.799  \\ 
                           & n=900 &  & 0.003 & 0.112 & 0.112 & 0.956 & 0.441  \\ 
\arrayrulecolor{black}  \hline
\end{tabular}
\end{center}
\end{table}

\begin{figure}[ht!]
	\centering	
	\includegraphics{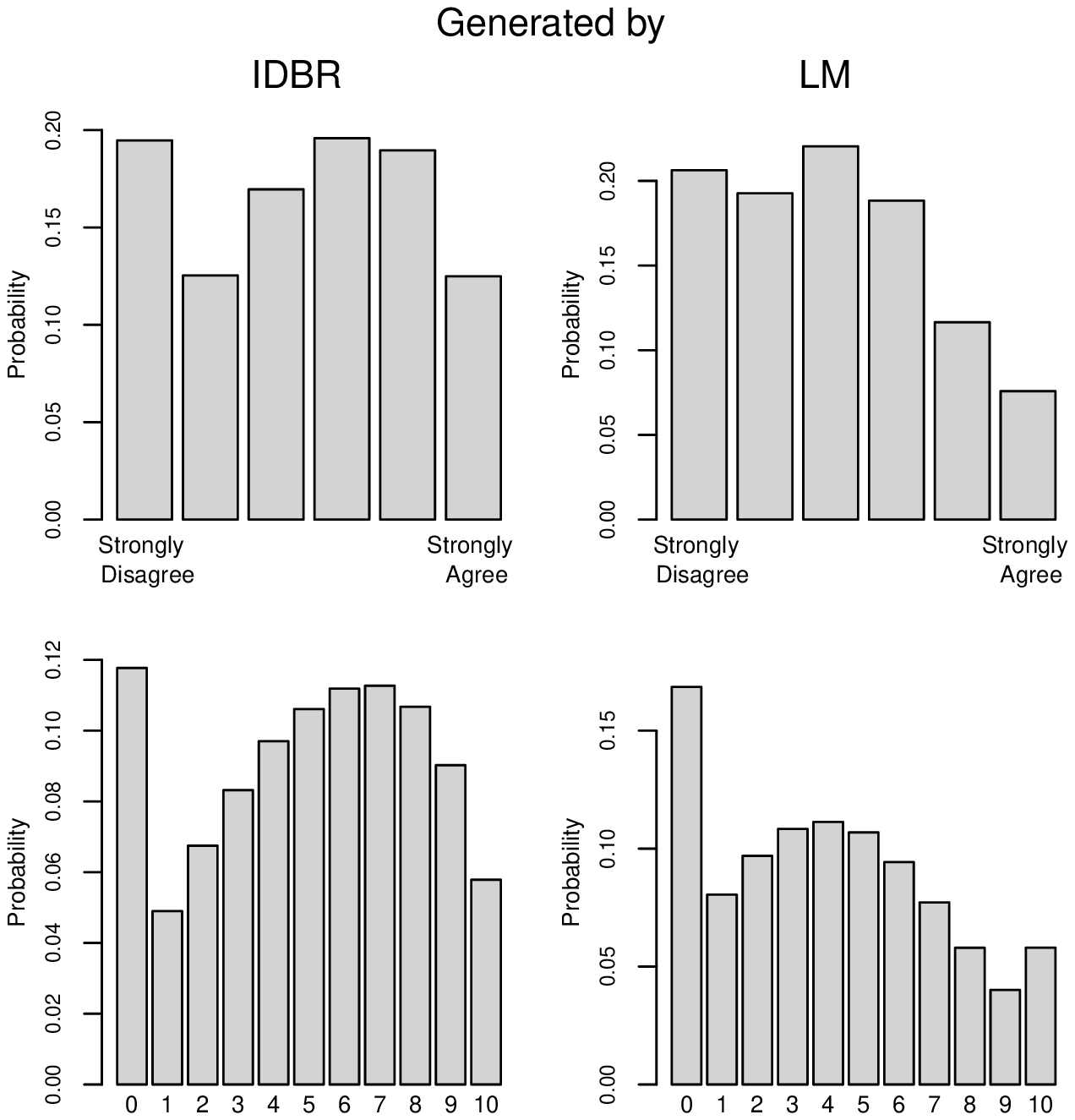} 
	\caption{Marginal distribution of the outcome across the four settings of 500 simulations}
	\label{fig:ExtBetaRegrSimuSampleSize6and11withLM}
\end{figure}

The results presented in table \ref{tab:ExtBetaRegrSimuSampleSize6} are based on 500 simulated datasets with a 6-level discrete scale outcome and seven covariates: variables V1 to V4 were generated from independent normal distributions with mean equals to 3 and variance equals to 1; dummies D1 to D3 were generated with probabilities of success equal to 0.5. The marginal distribution of the rates across the 500 simulations is illustrated on the top left part of figure \ref{fig:ExtBetaRegrSimuSampleSize6and11withLM} with purely illustrative labels. The bottom left part of the same figure shows the marginal distribution of the rates on the 11-level discrete scale with simulation results \\ \afterpage{\clearpage} \noindent gathered in table \ref{tab:ExtBetaRegrSimuSampleSize11}. %

As expected, tables \ref{tab:ExtBetaRegrSimuSampleSize6} and \ref{tab:ExtBetaRegrSimuSampleSize11} show that the bias, the standard deviation (Emp. SD), the root mean squared error (RMSE) and the length of HPD intervals decrease with sample size. The coverage of the credibility intervals are close to their nominal level even with moderate sample sizes. 


{\centering \section{Predictions} }
\label{sec:ExtBetaRegrPredictions}


To sample the predictive distribution of the response of subject $i$, one can rely on the MCMC sample for the regression parameters and the following algorithm. For iteration $l$ of the $L=3000$ draws after burn in and for observation or scenario $i$:  
\begin{enumerate}
	\item Draw $p_{i,l} \sim Uni\left[0,1\right]$ and set $\hat{y}_{i,l} = kh$ if $p_{i,l} \leq \left[1+\exp\left(-\mathbf{w}_i'\hat{\boldsymbol\gamma}_l\right)\right]^{-1}$ and proceed with step 2 otherwise;
	\item Draw $u_{i,l} \sim Be\left( \frac{\exp\left(-\mathbf{z}_i'\hat{\boldsymbol\theta}_l\right)}{1+\exp\left(-\mathbf{x}_i'\hat{\boldsymbol\beta}_l\right)} , \frac{\exp\left(-\mathbf{z}_i'\hat{\boldsymbol\theta}_l\right)}{1+\exp\left(\mathbf{x}_i'\hat{\boldsymbol\beta}_l\right)} \right)$ and round it to the upper bound of the proxy interval under the beta density: $\hat{y}_{i,l} = h ~ \lceil u_{i,l} / h \rceil$.
\end{enumerate}
 
The empirical distribution of $Y_i$ can be approximated using the empirical distribution of the Monte Carlo sample $\left\{\hat{y}_{i,l}\right\}_{l=1}^L$. A point prediction can be obtained from it by reporting its mode (say). Subset of values can also be constructed to predict $Y_i$. Forcing it to be an interval does not lead to good results in term of covering rate 
and interval length. Instead, we recommend to produce potentially disconnected regions that turn to have nice covering rates and smaller length. This method allows the $(1-\alpha)$ HPD to be disjoint if the inflated point has an estimated predictive probability greater than $\alpha$. In such cases, the $k^\mathrm{th}$ point of the scale should be part of the $(1-\alpha)$ HPD predictive region and added to the $(1-\alpha-\hat{\pi}_i)$ HPD predictive interval where $\hat{\pi}_i = \frac{1}{L} \sum_{l=1}^L{I\left(\hat{y}_{i,l}=kh\right)}$ is the estimated predictive probability of the inflated point.

In order to check the quality of this prediction scheme and to compare it with existing models, predictions were performed using five competing models: the multiple linear regression calculated using function \texttt{lm} in the R package \texttt{stats} \cite{RStats}, the continuous beta regression of \citeA{SimasBarretoSouza2010} using the function \texttt{betareg} in the R package \texttt{betareg} \cite{CribariNetoZeileis2010}, the proportional odds model calculated with function \texttt{clm} in the R package \texttt{ordinal} \cite{ROrdinal}, the multinomial logit model obtained using function \texttt{multinom} in the R package \texttt{nnet} \cite{RNnet} and our inflated discrete beta regression (IDBR) model. The continuous beta regression and the proportional odds model used here both link the location and the precision of the (underlying) distribution to covariates just as IDBR. Five hundreds datasets were generated in each simulation setting. Outcomes with 6 and 11 levels were simulated using either the inflated discrete beta regression (IDBR) or the linear regression (LM) models. Outcomes generated by linear regression were rounded to the nearest point of the discrete scale support. Predictions for dataset $(s+1)$ were made using parameter estimates from dataset $s$. Figure \ref{fig:ExtBetaRegrSimuSampleSize6and11withLM} illustrates the marginal distribution of the outcome in the four simulation settings.
\vskip 10pt

\begin{table}[ht!]
\begin{center}
\caption{Percentage of cases correctly predicted in the different simulation settings}
\label{tab:ExtBetaRegrSimuPredCorrect}
\begin{tabular}{cccccccc}
  \hline
Number & Generating & & \multicolumn{5}{c}{Estimating Model} \\
of Levels & Model & & \multicolumn{1}{c}{IDBR} & \multicolumn{1}{c}{\texttt{lm}} & \multicolumn{1}{c}{\texttt{betareg} }& \multicolumn{1}{c}{\texttt{clm}} & \multicolumn{1}{c}{\texttt{multinom}} \\ 
  \hline
\multirow{2}{*}{6} & IDBR & & 49.2\% & 36.8\% & 37.2\% & 47.0\% & 42.2\% \\ 
 & LM & & 44.1\% & 39.8\% & 41.5\% & 44.5\% & 43.4\% \\ 
  \hline
\multirow{2}{*}{11} & IDBR & & 32.4\% & 23.6\% & 23.3\% & 30.4\% & 26.9\% \\ 
 & LM & & 28.8\% & 21.2\% & 22.2\% & 29.8\% & 28.9\% \\ 
   \hline
\end{tabular}
\end{center}
\end{table}

Table \ref{tab:ExtBetaRegrSimuPredCorrect} reports the percentage of cases correctly predicted in the different simulation settings. Discrete response models (IDBR, \texttt{clm} and \texttt{multinom}) perform slightly better than the continuous ones (\texttt{lm} and \texttt{betareg}). Linear regression model (\texttt{lm}) is always the worse choice even when the outcome has been generated by it and simply rounded. The IDBR proves to be more accurate on datasets generated with inflation related to specific covariate pattern. When it is not, the IDBR is equivalent to its discrete competitor in term of proportion of correct predictions.
\vskip 10pt

\begin{table}[ht!]
\begin{center}
\caption{Interval coverage (cov.) and length in the different simulation settings}
\label{tab:ExtBetaRegrSimuPredHPD}
\begin{tabular}{cclcrrr}
  \hline
Number & Generating & Estimating & & \multicolumn{3}{c}{Interval (95\%)} \\ 
of Levels & Model & Model & & Cov. & Length$^1$ & Disjoint \\ 
  \hline
\multirow{10}{*}{6} & \multirow{5}{*}{IDBR} & IDBR & & 94.1\% & 0.550 & 34.6\% \\ 
& &  \texttt{lm} & & 96.5\% & 0.959 & \\ 
& &  \texttt{betareg} & & 97.2\% & 0.765 & \\ 
& &  \texttt{clm} & & 98.6\% & 0.766 & \\ 
& &  \texttt{multinom} & & 97.3\% & 0.749 & \\ 
  \cline{2-7}
& \multirow{5}{*}{LM} &  IDBR &  & 97.6\% & 0.567 & 0.1\% \\ 
& &  \texttt{lm} &  & 98.8\% & 0.738 & \\ 
& &  \texttt{betareg} &  & 98.7\% & 0.656 & \\ 
& &  \texttt{clm} &  & 99.2\% & 0.661 & \\ 
& &  \texttt{multinom} &  & 98.7\% & 0.618 & \\ 
  \hline
\multirow{10}{*}{11} & \multirow{5}{*}{IDBR} & IDBR & & 94.5\% & 0.488 & 37.1\% \\ 
& &  \texttt{lm} &   & 95.3\% & 0.827 & \\ 
& &  \texttt{betareg} &   & 95.6\% & 0.703 & \\ 
& &  \texttt{clm} &   & 97.2\% & 0.693 & \\ 
& &  \texttt{multinom} &   & 96.4\% & 0.686 & \\ 
  \cline{2-7}
& \multirow{5}{*}{LM} &  IDBR  &   & 96.7\% & 0.580 & 0.2\% \\ 
& &  \texttt{lm}  &   & 97.4\% & 0.711 & \\ 
& &  \texttt{betareg}   &  & 96.5\% & 0.639 & \\ 
& &  \texttt{clm}  &   & 98.7\% & 0.661 & \\ 
& &  \texttt{multinom}  &   & 97.8\% & 0.616 & \\ 
   \hline
\multicolumn{7}{l}{$^1$ Maximum interval length has been set to 1.} \\ 
\hline
\end{tabular}
\end{center}
\end{table}

Table \ref{tab:ExtBetaRegrSimuPredHPD} reports coverages and the length of the 95\% credible intervals. The coverages are generally close to their nominal values or tend to exceed them. The lengths have been computed by subtracting the upper and lower bound. When disjoint intervals (resulting from a separated inflation response) were produced using IDBR, the minimum distance between two consecutive points on the scale $h$ has been added to the main interval length to account for the addition of the inflation point. In order to make the interpretation comparable with different numbers of levels, all lengths have been rescaled to a maximum of 1 in the table. IDBR produces much narrower intervals than competing models even when the prediction region is disjoint: it indicates a better precision in predictions. When an inflation is introduced in the generating process, we get a large percentage of disjoint intervals (34.6 to 37.1\% in these simulations). This is not the case with datasets generated by the linear regression model even when there is an inflation due to rounding. Then, the presence of numerous disjoint intervals might be the symptom of subjects always selecting the inflated point among the proposed choices for particular covariate patterns and the relevance of the inflation submodel.



{\centering \section{Case Study}}
\label{sec:ExtBetaRegrCaseStudy}

\begin{figure}[ht!]
	\centering	
	\includegraphics{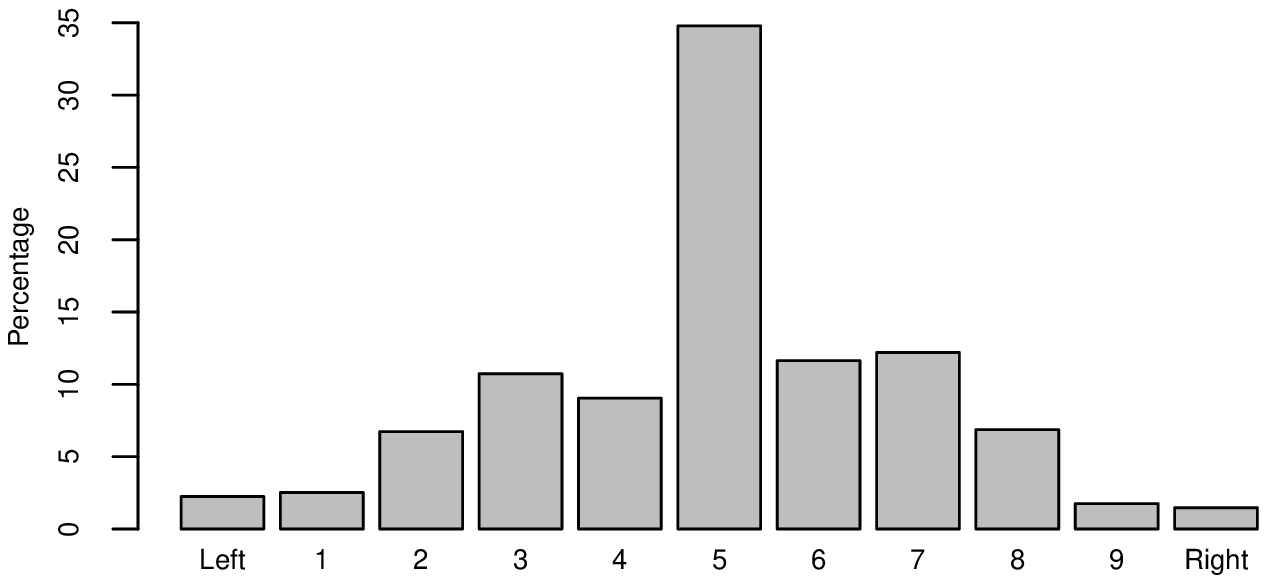} 
	\caption{Marginal shape of rates on the "left-right" political scale (B19) in Belgium (ESS Round 6)}
	\label{fig:ExtBetaRegrCaseStudy}
\end{figure}

The European Social Survey (ESS) is conducted every two years across thirty countries (in 2012) in Europe since 2001. Information about attitudes, beliefs and behaviors are collected during face to face interviews. Datasets are available on the Website: \href{http://www.europeansocialsurvey.org}{www.europeansocialsurvey.org}. We analyze the responses to question B19 of the ESS Round 6 \cite{ESS6}: \textit{In politics people sometimes talk of "left" and "right". Where would you place yourself on this scale, where 0 means the "left" and 10 means the "right"?} For simplicity, only the 1426 non-missing cases  from Belgium were used. Figure \ref{fig:ExtBetaRegrCaseStudy} illustrates the marginal shape of rates used in the model. Potential non-colinear covariates were selected and included in inflation, location and dispersion submodels. A backward selection procedure was used to simplify the model. The linearity of the effect has been checked for the ordered scale covariates. Table \ref{tab:ExtBetaRegrCaseStudy} shows the estimates of the final model where $p$ is the proportion of the MCMC sample after burnin with sign opposed to the one of the posterior median. 
\vskip 10pt

\begin{table}[ht!]
\small 
\begin{center}  
\caption{What does influence the placement on the "left-right" political scale (B19) in Belgium and who are those respondents choosing systematically the middle of the scale? (European Social Survey Round 6 \protect\cite{ESS6}, data from Belgium, 1426 non-missing cases only)}
\label{tab:ExtBetaRegrCaseStudy}
\begin{tabular}{llrrrr}
  \hline  
\multicolumn{2}{l}{\textbf{Covariates}} & \multicolumn{1}{c}{\multirow{2}{*}{\textbf{Estimate}}} & \multicolumn{2}{c}{\textbf{HPD(95\%)}} & \multicolumn{1}{c}{\multirow{2}{*}{\textbf{$p$}}} \\
(ESS Number) & Levels &  & Low & Up &  \\ 
  \hline  
 &  &  &  &  &  \\
  \hline  
\multicolumn{6}{l}{\textbf{Inflation}: Central level "5" $\phantom{e^{e^f}}$}  \\
\hline
(Intercept) &    & 0.087 & -0.580 & 0.812 & 0.413 \\ 
\arrayrulecolor{Gray} \hline
Gender & Male & 0 & . & . & .  \\
(F2) & Female & 0.478 & 0.117 & 0.815 & 0.004 \\ 
\hline
\multicolumn{2}{l}{Level of education (F15)}   & -0.164 & -0.244 & -0.073 & 0.000 \\ 
\hline
\multicolumn{2}{l}{Your place in society (D38)}   & -0.155 & -0.265 & -0.031 & 0.004 \\ 
 &  &  &  &  &  \\
\arrayrulecolor{black}  \hline  
\multicolumn{2}{l}{\textbf{Location} $\phantom{e^{e^f}}$}   &  &  &  &  \\ 
\hline
(Intercept) &    & -0.680 & -0.937 & -0.435 & 0.000  \\ 
\arrayrulecolor{Gray} \hline
Living Area & Big City & 0 & . & . & . \\
(F14) & Suburbs & 0.266 & 0.062 & 0.450 & 0.005 \\ 
 & Small City & 0.176 & 0.006 & 0.325 & 0.018 \\ 
 & Countryside & 0.216 & 0.068 & 0.351 & 0.003 \\ 
\hline
Gender & Male & 0 & . & . & . \\
(F2) & Female & -0.119 & -0.213 & -0.020 & 0.008 \\ 
\hline
\multicolumn{2}{l}{Household's total net income (F41)}   & 0.023 & 0.001 & 0.046 & 0.018 \\ 
\hline
\multicolumn{2}{l}{Your place in society (D38)}   & 0.068 & 0.029 & 0.104 & 0.000 \\ 
 &  &  &  &  &  \\
\arrayrulecolor{black}  \hline  
\multicolumn{2}{l}{\textbf{Dispersion} $\phantom{e^{e^f}}$}   &  &  &  &  \\ 
\hline
(Intercept) &    & -1.957 & -2.291 & -1.586 & 0.000 \\ 
\arrayrulecolor{Gray} \hline
\multicolumn{2}{l}{Age (F3)}   & 0.009 & 0.004 & 0.013 & 0.000 \\ 
\hline
\multicolumn{2}{l}{Level of education (F15)}   & -0.053 & -0.099 & -0.006 & 0.013 \\ 
\hline  
\multicolumn{2}{l}{Feeling about household's income}   & \multirow{2}{*}{0.162} & \multirow{2}{*}{0.060} & \multirow{2}{*}{0.271} &  \multirow{2}{*}{0.003} \\ 
\multicolumn{2}{l}{(Low=Confortable, F42)}   &  &  &  &   \\ 
\arrayrulecolor{black}   \hline  
\end{tabular}  
\end{center}  
\end{table}  

Rates in figure \ref{fig:ExtBetaRegrCaseStudy} bring the inflation clearly out. About 35\% of the respondents choose the central level of the scale, the only one which is neither on the left nor on the right. The hypothesis of the inflated discrete beta regression (IDBR) model is that those respondents might be divided in two groups: those who systematically choose the middle of the scale, saying \textit{I don't care about politics} or \textit{I don't identify myself in the left-right opposition}; and those who choose that point but might have choosen the neighbouring alternatives. This distinction is meaningful in Belgium since some political parties even unequivocally position themselves in the centre of the left-right scale using their name (e.g.: CDH=\textit{Centre démocrate humaniste}). The model suggests that three covariates are particularly pertinent in modeling the invariant central-level choosers: Being a women multiply by 1.61 ($= \exp(0.478)$) the odds of choosing systematically the middle of the scale (when all other things remains equals). The higher the level of education, the lower the probability of not positioning on the left-right scale: for each level of the International Standard Classification of Education (ISCED), the odds is divided by 1.18 ($= 1/\exp(-0.164)$). Finally, responses to question D38 ("\textit{There are people who tend to be towards the top of our society and people who tend to be towards the bottom. On a scale that runs from top (=10) to bottom (=0), where would you place yourself nowadays?}") also influence the probability of systematically choosing the middle of the scale: the higher you place yourself in the society, the lower the probability of choosing systematically the middle of the scale.

Looking at the whole scale now, IDBR suggests that four covariates are pertinent to understand the location of a respondent on the scale: their living area, their gender, their income and, again, where they place themselves in the society. People living in big cities tend to place themselves more on the left whereas people from suburbs tend to favor righter choices. This corresponds to trends observed after each election in Belgium. Women are positioned more to the left than men as observed for years now. The household's total net income and the place in society both influence the location on the scale with the better off more often found on the right. The latter two covariates can be included together in the model since they are not clearly correlated (Pearson $r=0.27$) nor colinear ($VIF = 1.13$).

For a given location on the scale, what does influence the dispersion of the placement on the "left-right" political scale in Belgium? IDBR suggests that two covariates increase the dispersion of the rates and one decreases it. The age of the respondent increases the dispersion of the responses, suggesting that older people are more dispersed on the scale than younger ones. People who have the feeling to live comfortably on their present income (F42=1) are generally less dispersed on the left-right scale than those who find it very difficult to live on present income (F42=4). This conclusion corresponds to the fact that extreme left and right voters often justify their vote as the only available response to their discomfort or frustration. Finally, the higher the level of education, the lower the dispersion of the responses.



{\centering \section{Discussion} }
\label{sec:ExtBetaRegrDiscussion}


In this paper, we address four critical characteristics of Likert and rating scale outcomes that have never been taken into account simultaneously by existing models. Discreteness, boundedness and potential skewness are handled by the discrete beta regression (DBR) model described in section 2.
The ability to identify likely invariant choosers of a particular level in the scale is one of the relevant features of the inflated discrete beta regression (IDBR) model described in section 3.
The excellent statistical properties of the latter tool suggest that it is a valuable alternative to the existing (and less comprehensive) methods to model selection behaviors and to make predictions.

The DBR and IDBR models produce valuable outputs in both interpretation and prediction processes. Identifying the covariates influencing the location of the outcome is a classical model output, but understanding what determines unanimity or indecisiveness in subgroups of respondents might be determinant when for example marketers launch a campaign or doctors evaluate patient's pain. Distinguishing between invariant and less resolute choosers of a particular alternative in a Likert scale could for example save money to announcers by not sending ads to likely invariable non buyers. Compared with competing models, the IDBR model produces discrete predictions with high proportion of cases correctly predicted and the narrowest credible intervals.

Extension of IDBR to multiple inflations (e.g.: both bounds are inflated) is straightforward but has not been tested yet. Adaptation of the IDBR to hierarchical experiment such as conjoint analysis or psychological measurements would have numerous applications. It will be our next step of research.



\bibliographystyle{apacite} 
\bibliography{IDBRcTaverneBiblio}

\begin{thebibliography}{}

\bibitem[\protect\citeauthoryear{%
Atchade%
\ \BBA{} Rosenthal%
}{%
Atchade%
\ \BBA{} Rosenthal%
}{%
{\protect\APACyear{2005}}%
}]{%
AtchadeRosenthal2005}%
\APACinsertmetastar{%
AtchadeRosenthal2005}%
Atchade, Y\BPBI F.%
\BCBT{}\ \BBA{} Rosenthal, J\BPBI S.%
%
\unskip\
\newblock
\APACrefYearMonthDay{2005}{}{}.
\newblock
\BBOQ{}\APACrefatitle{On adaptive Markov chain Monte Carlo algorithms}{On
  adaptive markov chain monte carlo algorithms}.\BBCQ{}
\newblock
\APACjournalVolNumPages{Bernoulli}{11}{5}{815-828}.
\PrintBackRefs{\CurrentBib}

\bibitem[\protect\citeauthoryear{%
Branscum%
, Johnson%
\BCBL{}\ \BBA{} Thurmond%
}{%
Branscum%
\ \protect\BOthers{.}}{%
{\protect\APACyear{2007}}%
}]{%
BranscumJohnson2007}%
\APACinsertmetastar{%
BranscumJohnson2007}%
Branscum, A\BPBI J.%
, Johnson, W\BPBI O.%
\BCBL{}\ \BBA{} Thurmond, M\BPBI C.%
%
\unskip\
\newblock
\APACrefYearMonthDay{2007}{}{}.
\newblock
\BBOQ{}\APACrefatitle{Bayesian beta regression - Applications to household
  expenditure data and genetic distance between foot-and-mouth disease
  viruses}{Bayesian beta regression - applications to household expenditure
  data and genetic distance between foot-and-mouth disease viruses}.\BBCQ{}
\newblock
\APACjournalVolNumPages{Australian \& New Zealand Journal of
  Statistics}{49}{3}{287-301}.
\PrintBackRefs{\CurrentBib}

\bibitem[\protect\citeauthoryear{%
Brehm%
\ \BBA{} Gates%
}{%
Brehm%
\ \BBA{} Gates%
}{%
{\protect\APACyear{1993}}%
}]{%
BrehmGates1993}%
\APACinsertmetastar{%
BrehmGates1993}%
Brehm, J.%
\BCBT{}\ \BBA{} Gates, S.%
%
\unskip\
\newblock
\APACrefYearMonthDay{1993}{}{}.
\newblock
\BBOQ{}\APACrefatitle{Donut Shops and Speed Traps: Evaluating Models of
  Supervision on Police Behavior}{Donut shops and speed traps: Evaluating
  models of supervision on police behavior}.\BBCQ{}
\newblock
\APACjournalVolNumPages{American Journal of Political
  Science}{37}{2}{555--581}.
\PrintBackRefs{\CurrentBib}

\bibitem[\protect\citeauthoryear{%
Christensen%
}{%
Christensen%
}{%
{\protect\APACyear{2013}}%
}]{%
ROrdinal}%
\APACinsertmetastar{%
ROrdinal}%
Christensen, R\BPBI H\BPBI B.%
%
\unskip\
\newblock
\APACrefYearMonthDay{2013}{}{}.
\newblock
\APACrefbtitle{ordinal---Regression Models for Ordinal
  Data.}{ordinal---regression models for ordinal data.}
\newblock
\APACrefnote{R package version 2013.9-30}
\PrintBackRefs{\CurrentBib}

\bibitem[\protect\citeauthoryear{%
ESS%
}{%
ESS%
}{%
{\protect\APACyear{2012}}%
}]{%
ESS6}%
\APACinsertmetastar{%
ESS6}%
ESS.%
%
\unskip\
\newblock
\APACrefYearMonthDay{2012}{}{}.
\newblock
\APACrefbtitle{European Social Survey - Round 6 Data.}{European social survey -
  round 6 data.}
\newblock
\APAChowpublished{Data file edition 1.2. Norwegian Social Science Data
  Services, Norway - Data Archive and distributor of ESS data.}
\PrintBackRefs{\CurrentBib}

\bibitem[\protect\citeauthoryear{%
Ferrari%
\ \BBA{} Cribari-Neto%
}{%
Ferrari%
\ \BBA{} Cribari-Neto%
}{%
{\protect\APACyear{2004}}%
}]{%
FerrariCribariNeto2004}%
\APACinsertmetastar{%
FerrariCribariNeto2004}%
Ferrari, S.%
\BCBT{}\ \BBA{} Cribari-Neto, F.%
%
\unskip\
\newblock
\APACrefYearMonthDay{2004}{}{}.
\newblock
\BBOQ{}\APACrefatitle{Beta Regression for Modelling Rates and Proportions}{Beta
  regression for modelling rates and proportions}.\BBCQ{}
\newblock
\APACjournalVolNumPages{Journal of Applied Statistics}{31}{7}{799-815}.
\PrintBackRefs{\CurrentBib}

\bibitem[\protect\citeauthoryear{%
Gelman%
\ \BBA{} Rubin%
}{%
Gelman%
\ \BBA{} Rubin%
}{%
{\protect\APACyear{1992}}%
}]{%
GelmanRubin1992}%
\APACinsertmetastar{%
GelmanRubin1992}%
Gelman, A.%
\BCBT{}\ \BBA{} Rubin, D\BPBI B.%
%
\unskip\
\newblock
\APACrefYearMonthDay{1992}{}{}.
\newblock
\BBOQ{}\APACrefatitle{Inference from iterative simulation using multiple
  sequences}{Inference from iterative simulation using multiple
  sequences}.\BBCQ{}
\newblock
\APACjournalVolNumPages{Statistical Science}{7}{}{457-511}.
\PrintBackRefs{\CurrentBib}

\bibitem[\protect\citeauthoryear{%
Lambert%
}{%
Lambert%
}{%
{\protect\APACyear{1992}}%
}]{%
Lambert1992}%
\APACinsertmetastar{%
Lambert1992}%
Lambert, D.%
%
\unskip\
\newblock
\APACrefYearMonthDay{1992}{{\APACmonth{02}}}{}.
\newblock
\BBOQ{}\APACrefatitle{Zero-inflated Poisson Regression, with an Application to
  Defects in Manufacturing}{Zero-inflated poisson regression, with an
  application to defects in manufacturing}.\BBCQ{}
\newblock
\APACjournalVolNumPages{Technometrics}{34}{1}{1--14}.
\PrintBackRefs{\CurrentBib}

\bibitem[\protect\citeauthoryear{%
Ospina%
\ \BBA{} Ferrari%
}{%
Ospina%
\ \BBA{} Ferrari%
}{%
{\protect\APACyear{2010}}%
}]{%
OspinaFerrari2010}%
\APACinsertmetastar{%
OspinaFerrari2010}%
Ospina, R.%
\BCBT{}\ \BBA{} Ferrari, S\BPBI L.%
%
\unskip\
\newblock
\APACrefYearMonthDay{2010}{}{}.
\newblock
\BBOQ{}\APACrefatitle{Inflated beta distributions}{Inflated beta
  distributions}.\BBCQ{}
\newblock
\APACjournalVolNumPages{Statistical Papers}{51}{1}{111-126}.
\PrintBackRefs{\CurrentBib}

\bibitem[\protect\citeauthoryear{%
Ospina%
\ \BBA{} Ferrari%
}{%
Ospina%
\ \BBA{} Ferrari%
}{%
{\protect\APACyear{2012}}%
}]{%
OspinaFerrari2012}%
\APACinsertmetastar{%
OspinaFerrari2012}%
Ospina, R.%
\BCBT{}\ \BBA{} Ferrari, S\BPBI L.%
%
\unskip\
\newblock
\APACrefYearMonthDay{2012}{}{}.
\newblock
\BBOQ{}\APACrefatitle{A general class of zero-or-one inflated beta regression
  models}{A general class of zero-or-one inflated beta regression
  models}.\BBCQ{}
\newblock
\APACjournalVolNumPages{Computational Statistics \& Data
  Analysis}{56}{6}{1609-1623}.
\PrintBackRefs{\CurrentBib}

\bibitem[\protect\citeauthoryear{%
Paolino%
}{%
Paolino%
}{%
{\protect\APACyear{2001}}%
}]{%
paolino2001}%
\APACinsertmetastar{%
paolino2001}%
Paolino, P.%
%
\unskip\
\newblock
\APACrefYearMonthDay{2001}{}{}.
\newblock
\BBOQ{}\APACrefatitle{Maximum Likelihood Estimation of Models with
  Beta-Distributed Dependent Variables}{Maximum likelihood estimation of models
  with beta-distributed dependent variables}.\BBCQ{}
\newblock
\APACjournalVolNumPages{Political Analysis}{9}{4}{325--346}.
\PrintBackRefs{\CurrentBib}

\bibitem[\protect\citeauthoryear{%
Plummer%
, Best%
, Cowles%
\BCBL{}\ \BBA{} Vines%
}{%
Plummer%
\ \protect\BOthers{.}}{%
{\protect\APACyear{2006}}%
}]{%
PlummerBest2006}%
\APACinsertmetastar{%
PlummerBest2006}%
Plummer, M.%
, Best, N.%
, Cowles, K.%
\BCBL{}\ \BBA{} Vines, K.%
%
\unskip\
\newblock
\APACrefYearMonthDay{2006}{}{}.
\newblock
\BBOQ{}\APACrefatitle{CODA: Convergence Diagnosis and Output Analysis for
  MCMC}{Coda: Convergence diagnosis and output analysis for mcmc}.\BBCQ{}
\newblock
\APACjournalVolNumPages{R News}{6}{1}{7--11}.
\PrintBackRefs{\CurrentBib}

\bibitem[\protect\citeauthoryear{%
{R Core Team}%
}{%
{R Core Team}%
}{%
{\protect\APACyear{2012}}%
}]{%
RStats}%
\APACinsertmetastar{%
RStats}%
{R Core Team}.%
%
\unskip\
\newblock
\APACrefYearMonthDay{2012}{}{}.
\newblock
\BBOQ{}\APACrefatitle{R: A Language and Environment for Statistical
  Computing}{R: A language and environment for statistical computing}\BBCQ{}\
  [\bibcomputersoftwaremanual].
\newblock
\APACaddressPublisher{Vienna, Austria}{}.
\PrintBackRefs{\CurrentBib}

\bibitem[\protect\citeauthoryear{%
Simas%
, Barreto-Souza%
\BCBL{}\ \BBA{} Rocha%
}{%
Simas%
\ \protect\BOthers{.}}{%
{\protect\APACyear{2010}}%
}]{%
SimasBarretoSouza2010}%
\APACinsertmetastar{%
SimasBarretoSouza2010}%
Simas, A\BPBI B.%
, Barreto-Souza, W.%
\BCBL{}\ \BBA{} Rocha, A\BPBI V.%
%
\unskip\
\newblock
\APACrefYearMonthDay{2010}{February}{}.
\newblock
\BBOQ{}\APACrefatitle{Improved estimators for a general class of beta
  regression models}{Improved estimators for a general class of beta regression
  models}.\BBCQ{}
\newblock
\APACjournalVolNumPages{Computational Statistics \& Data
  Analysis}{54}{2}{348-366}.
\PrintBackRefs{\CurrentBib}

\bibitem[\protect\citeauthoryear{%
Venables%
\ \BBA{} Ripley%
}{%
Venables%
\ \BBA{} Ripley%
}{%
{\protect\APACyear{2002}}%
}]{%
RNnet}%
\APACinsertmetastar{%
RNnet}%
Venables, W\BPBI N.%
\BCBT{}\ \BBA{} Ripley, B\BPBI D.%
%
\unskip\
\newblock
\APACrefYear{2002}.
\newblock
\APACrefbtitle{Modern Applied Statistics with S}{Modern applied statistics with
  s}\ (\PrintOrdinal{Fourth}\ \BEd).
\newblock
\APACaddressPublisher{New York}{Springer}.
\PrintBackRefs{\CurrentBib}

\bibitem[\protect\citeauthoryear{%
Wieczorek%
\ \BBA{} Hawala%
}{%
Wieczorek%
\ \BBA{} Hawala%
}{%
{\protect\APACyear{2011}}%
}]{%
WieczorekHawala2011}%
\APACinsertmetastar{%
WieczorekHawala2011}%
Wieczorek, J.%
\BCBT{}\ \BBA{} Hawala, S.%
%
\unskip\
\newblock
\APACrefYearMonthDay{2011}{}{}.
\newblock
\BBOQ{}\APACrefatitle{A Bayesian Zero-One Inflated Beta Model for Estimating
  Poverty in U.S. Counties}{A bayesian zero-one inflated beta model for
  estimating poverty in u.s. counties}.\BBCQ{}
\newblock
\BIn{} \APACrefbtitle{Proceedings of the American Statistical Association,
  Section on Survey Research Methods, Alexandria, VA.}{Proceedings of the
  american statistical association, section on survey research methods,
  alexandria, va.}
\PrintBackRefs{\CurrentBib}

\bibitem[\protect\citeauthoryear{%
Zeileis%
\ \BBA{} Cribari-Neto%
}{%
Zeileis%
\ \BBA{} Cribari-Neto%
}{%
{\protect\APACyear{2010}}%
}]{%
CribariNetoZeileis2010}%
\APACinsertmetastar{%
CribariNetoZeileis2010}%
Zeileis, A.%
\BCBT{}\ \BBA{} Cribari-Neto, F.%
%
\unskip\
\newblock
\APACrefYearMonthDay{2010}{}{}.
\newblock
\BBOQ{}\APACrefatitle{Beta Regression in R}{Beta regression in r}.\BBCQ{}
\newblock
\APACjournalVolNumPages{Journal of Statistical Software}{34}{2}{1-24}.
\PrintBackRefs{\CurrentBib}

\end{thebibliography}


\end{document}